\documentstyle[12pt,psfig]{article}
\let\section=\subsection     \let\subsection=\subsubsection               
                          
\begin{document}
\begin{center}
   {\large \bf A NEW MECHANISM FOR}\\[2mm]
   {\large \bf J/$\psi$ SUPPRESSION IN NUCLEAR COLLISIONS}\\[5mm]
   ALBERTO POLLERI\\[5mm]
   {\small \it Institut f\"ur Theoretische Physik der Universit\"at, \\
     Philosophenweg 19, D-69120 Heidelberg, Germany. \\
     e-mail: polleri@tphys.uni-heidelberg.de \\[8mm] }
\end{center}

\begin{abstract}\noindent
We present new results based on an improved version of the Glauber model,
used to describe the nuclear absorption of J/$\psi$ in reactions between
nuclei at high 
energy. The excitation of the colour degrees of freedom of nucleons, due to
collisions mediated by one-gluon exchange, is taken into account. It is found
that the proposed mechanism leads to larger nuclear absorption of J/$\psi$
than previously considered.\\
\end{abstract}

The origin of the anomalous behaviour of the J/$\psi$ cross section measured
in Pb+Pb collisions \cite{NA50} is still not understood and competing
interpretations have been proposed \cite{GHV}. In order to resolve this 
important and controversial issue, more detailed calculations are demanded.
Providing further insight in this difficult problem, preliminary results, 
obtained with a new formulation of the Glauber approach to the early stage of
a nuclear collision, are hereby presented. 

Preparing the ground for the discussion, it is useful to recall the 
Glauber model expression for the cross section in a A+B collision at
impact parameter $\vec{b}$. Denoting a generic colour
singlet $c\bar{c}$ state by $\psi$, one has
\begin{equation}
\frac{d^2\sigma_\psi^{A B}}{d^2\vec{b}}(\vec{b}) = \sigma_\psi^{N N} 
\!\!\int\! d^2\vec{s}\ dz_A\, dz_B \ \rho_A(z_A,\vec{s}) 
\, \rho_B(z_B,\vec{b}-\vec{s})\ S^{abs}(z_A,z_B,\vec{b},\vec{s})\,,
\label{NUCCROSS}
\end{equation}
where $\rho_A$ and $\rho_B$ are the densities of the colliding nuclei and
\begin{equation}
S^{abs}(z_A,z_B,\vec{b},\vec{s})\! = \exp\left[ - \sigma^{abs}_{\psi N} 
\left(\int_{- \infty}^{z_A}\!\! dz'_A\, \rho_A(z'_A,\vec{s}) \! + \!\!
\int_{z_B}^{+ \infty}\!\!\! dz'_B\, \rho_B(z'_B,\vec{b}-\vec{s})\!\right)\! \right]
\label{EXPON}
\end{equation}
is the nuclear suppression factor, which contains the effective absorption
cross section $\sigma^{abs}_{\psi N} $ for $\psi$-nucleon scattering.

Provided that $\sigma^{abs}_{\psi N} = 7.3$ mb, the above expressions can 
account for p+A and S+Pb data, but not for the Pb+Pb measurements.
On the other hand, the usual version of the Glauber model must be improved,
since in eqs.$\,$(\ref{NUCCROSS}) and (\ref{EXPON}) one assumes that the 
nucleons which absorb the produced $\psi$ are 
in their ground state. This cannot be correct. In fact, before
encountering the produced meson, they scatter several times while
the nuclei stream through each other. In doing so they leave the ground state.
A first attempt to address this problem was recently made, by including in
the calculation the effect of the cloud of prompt gluons around nucleons 
\cite{HK}. On the other hand, nucleon themselves were still treated as 
non-interacting.
Since the centre of mass energy of each NN collision is $\simeq 17$ GeV at the
SPS, it is reasonable to ascribe the main contribution to the inelastic cross 
sections to one-gluon exchange processes, as done, for example, in the
Dual Parton Model \cite{DPM}. This has an important
consequence: two colour-singlet nucleons jump into octet
states after the elementary collision and therefore become {\it coloured}. 
In same way one also allows the possibility that some 
`nucleons' are in a decuplet state. From now on the quotes will be dropped when
referring to a coloured nucleon.

Having realized that nucleons become coloured, soon after enough rescattering
has taken place, it is necessary to establish whether this fact has any 
effect at all on $\psi$ absorption. In other words one must quantify the 
differences in the inelastic cross sections for $\psi\:$N scattering due
to the various colour states of the nucleon. This can be achieved within the
Low-Nussinov model \cite{LN}, first evaluating the elastic cross section for 
a meson scattering off a nucleon, and then making use of the optical theorem, 
which relates the forward elastic amplitude to the total cross section.
The model consists in the exchange of two gluons characterised by a 
phenomenological mass $\mu_G \simeq 140$ MeV, which effectively mimics 
confinement. Neglecting the elastic contribution compared to the inelastic 
one, one can show that
\begin{equation}
\sigma^{abs}_{\psi N}\, = \int d^2\rho 
\, \left|\Phi_\psi(\rho)\right|^2 \sigma_a(\rho)\,,
\label{TOTAL}
\end{equation}
where $\Phi_\psi(\rho)$ is the transverse part of the $\psi$ meson wave 
function, while
\begin{equation}
\sigma_a(\rho)\, = \frac{16}{3}\, \alpha_s^2\, \int d^2k 
\ \frac{1}{(k^2 + \mu_G^2)^2}
\, \left[1 - a\,F_N(3k^2)\right] 
\, \left(1 - e^{i \,\vec{k}\cdot \vec{\rho}}\right)
\label{DIP}
\end{equation}
is the colour dipole-nucleon cross section.
The function $F_N$ is the two-quark form factor of the nucleon, usually 
identified with the charge form factor. 
The coefficient $a$ in front of the form factor specifies the colour state of
the nucleon. For the singlet state one has $a = 1$ and the expression reduces 
to the usual one, exhibiting the partial cancellation of amplitudes, 
implying $\sigma(\rho) \rightarrow 0$ for $\rho \rightarrow 0$ (Colour 
transparency). A detailed calculation shows that $a = 1/4$ for an 
octet nucleon while $a = - 1/2$ for the decuplet state.
\begin{figure}[t]
\centerline{\psfig{figure=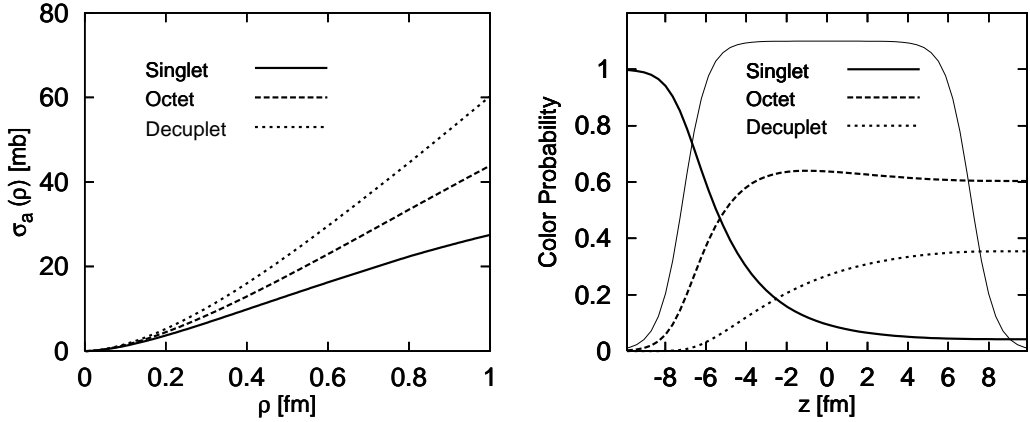,width=14cm,angle=-90}} 
\protect\caption{Left - Dipole cross section for different colour states. 
Right - Evolution of the colour probabilities $P_S$, $P_O$, $P_D$, along $z$
and with $\vec{b} = 0$. Also shown is the profile of a Pb nucleus.}
\label{ONE}
\end{figure}
The dipole cross section can be computed analytically in terms of modified
Bessel functions. One can fix the value of the strong coupling in order
to reproduce $\pi\,$N cross section of $\simeq 23$ mb, therefore taking
$\alpha_s \simeq 0.65$.
Successively, the $\psi\,$N cross sections can be obtained using 
eqs.$\,$(\ref{TOTAL}) and (\ref{DIP}) without additional parameter fixing.
The results of the calculations are illustrated in the left side of 
Figure \ref{ONE}, which shows the dipole cross section for the three colour
states of the nucleon. It largely increases when the dipole scatters off a
coloured object. This new result is important and could not have been guessed
prior to calculation.
One must now evaluate the corresponding absorption cross sections.
To do so it is useful to first rewrite eq.$\,$(\ref{DIP}) in the form 
\begin{equation}
\sigma_a(\rho) = \sigma(\rho) \left [\, 1 + (1 - a)\,\Delta(\rho) \,\right]\,,
\end{equation}
where the function $\Delta(\rho)$ was found to be approximately constant and 
amounts to $\simeq 0.46$. This shows that the colour cross sections have a 
simple $\rho$-dependence and scale with respect to the singlet one. 
Then, using eq.$\,$(\ref{TOTAL}), one obtains the absorption cross sections
for $\psi$ as
\begin{equation}
\sigma^{abs}_{\psi N_j} = \sigma^{abs}_{\psi N} \ \times \
\left\{{
\begin{tabular}{l l} 
1     &  Singlet    \\ 
1.35   &  Octet      \\ 
1.7   &  Decuplet   \\ 
\end{tabular}}\right.\,.
\label{COLCROSS}
\end{equation}
A large increase with respect to the singlet value is found, suggesting that 
the newly proposed absorption mechanism is indeed important. Using a Gaussian 
parametrisation for the $\psi$ wave function, such that the root mean
squared radius is $\sim 0.2$ fm, one obtains $\sigma^{abs}_{\psi N} \simeq 6$
mb. On the other hand, at this preliminary stage of the calculation, the
effect of the feeding into J/$\psi$ from $\psi'$ and $\chi_c$ states is
neglected. To compare with the conventional Glauber model, it is therefore
preferable to use the effective value used to reproduce the p+A data, 
therefore setting $\sigma^{abs}_{\psi N} = 7.3$ mb and scaling the colour cross
section accordingly by means of eq.$\,$(\ref{COLCROSS}). 

To correct eqs.$\,$(\ref{NUCCROSS}) and (\ref{EXPON}),
one must now understand how the colour state of a nucleon evolves while 
passing through a nucleus. This is a non-trivial problem which can be
formulated by means of a master equation. Its content is to express the 
evolution of the colour probabilities $P_S$, $P_O$ and $P_D$,
due to subsequent scatterings. The framework is again
that of the Glauber model of multiple collisions. What one finds is the
solution \cite{BOR}
\begin{eqnarray}
P_S(z,\vec{b}) & = & \frac{1}{27} \ + \ \frac{20}{27}\ F(z,\vec{b}) \ + 
\ \frac{6}{27}\ G(z,\vec{b})\,, \label{MAST1}\\
P_O(z,\vec{b}) & = & \frac{16}{27} \ - \ \frac{40}{27}\ F(z,\vec{b}) \ + 
\ \frac{24}{27}\ G(z,\vec{b})\,, \label{MAST2}\\
P_D(z,\vec{b}) & = & \frac{10}{27} \ + \ \frac{20}{27}\ F(z,\vec{b}) \ - 
\ \frac{30}{27}\ G(z,\vec{b})\,,\label{MAST3}
\end{eqnarray}
where
\begin{equation}
F(z,\vec{b}) = \exp\left(-X(z,\vec{b})\right)\,,\ \ \ \ 
G(z,\vec{b}) = \exp\left(-\mbox{\large$\frac{2}{3}$}X(z,\vec{b})\right)
\end{equation}
and
\begin{equation}
X(z,\vec{b}) = \mbox{\large$\frac{9}{8}$}\ \sigma^{in}_{NN} 
\int_{- \infty}^z \!\! dz'\ \rho(z',\vec{b})\,, \ \ \ \ \mbox{with}\ \ \ \ 
\sigma^{in}_{NN} = 30\, \mbox{mb}\,.
\end{equation}
One notices several properties of the found probabilities. First of all, the
limit $z \rightarrow -\infty$ implies $X \rightarrow 0$ and $F,G \rightarrow 
1$. This means that 
$P_S \rightarrow 1$ and $P_O,P_D \rightarrow 0$. In other words the nucleon,
before entering the nucleus, is in singlet state as it should be. Moreover,
$P_S + P_O + P_D = 1$ for any $z$, therefore probability is always conserved.
Finally, if $z \rightarrow +\infty$ and if the nucleus is large enough,
one has $X \gg 1$ and $F,G \ll 1$. This implies that if enough scatterings 
take place, the colour probabilities reach the statistical limit, given by the 
first coefficients of eqs.$\,$(\ref{MAST1}), (\ref{MAST2}) and (\ref{MAST3}).
The $z$-dependence of the colour probabilities is illustrated in the right
side of Figure \ref{ONE}, together with the longitudinal profile of a Pb
nucleus at $\vec{b} = 0$. Soon after the nucleon has penetrated the nuclear 
profile, a process that involves $\sim 4$ fm, the statistical limit is 
reached. About $2/3$ of the nucleons are in octet state while $1/3$
are in decuplet. The amount of singlet is negligible.
\begin{figure}[t]
\centerline{\psfig{figure=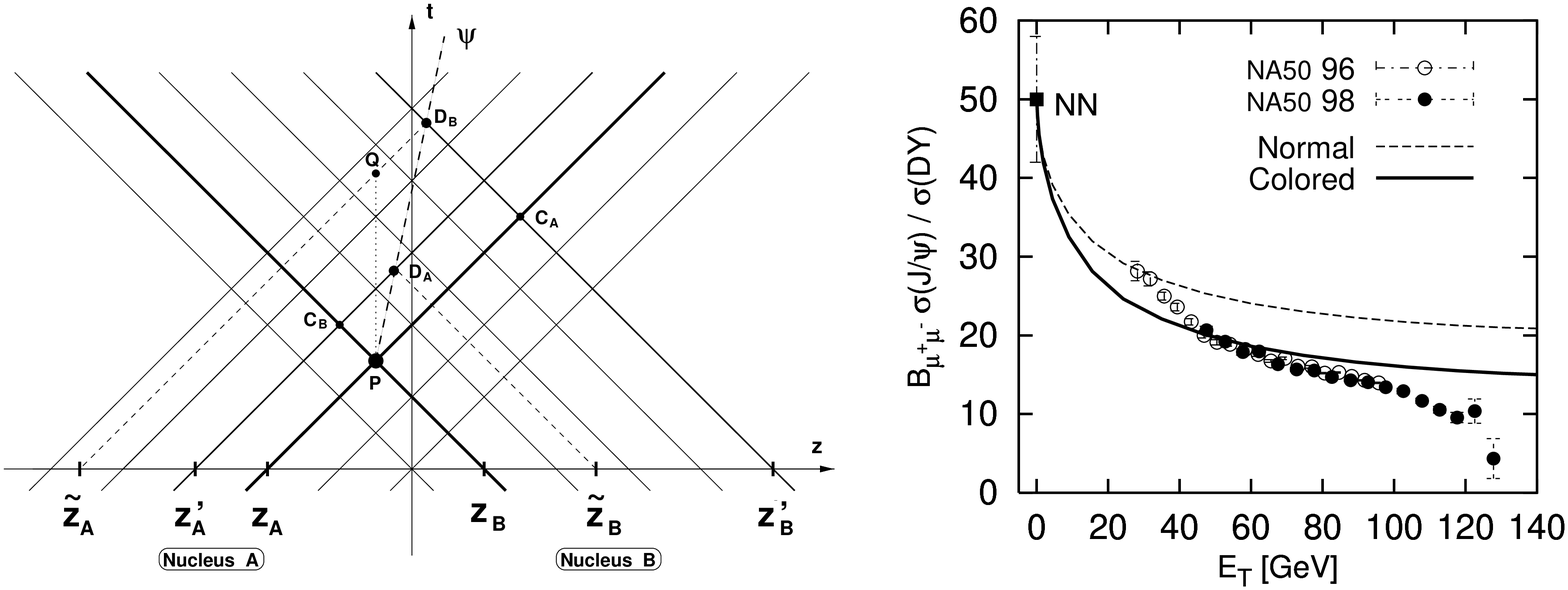,width=14cm}} 
\protect\caption{Left - Light cone representation of the collisions taking
place at impact parameters $(\vec{b},\vec{s})$. Right - Calculations of the
J/$\psi$ cross section
compared with the NA50 data. The curves correspond to normal nuclear 
absorption (dashed) and to absorption by coloured nucleons (full).}
\label{TWO}
\end{figure}

It is now possible to describe how to perform the calculation of the $\psi$ 
cross section in A+B collisions. One must modify eqs.$\,$(\ref{NUCCROSS}) and 
(\ref{EXPON}) to account for the previously discussed colour absorption cross 
sections and probabilities. It is necessary to replace eq.$\,$(\ref{EXPON})
with $S^{abs} = \exp\left[- (f_A + f_B)\right]$, where
\begin{eqnarray}
f_A(z_A,z_B,\vec{b},\vec{s}) & = & \int_{- \infty}^{z_A} \!\!dz'_A
\ \Sigma_B(\tilde{z}_B,\vec{b}-\vec{s})\ \rho_A(z'_A,\vec{s})\,,\\
f_B(z_A,z_B,\vec{b},\vec{s}) & = & \int_{z_B}^{+ \infty} \!\!dz'_B
\ \Sigma_A(\tilde{z}_A,\vec{s})\ \rho_B(z'_B,\vec{b}-\vec{s})\,
\end{eqnarray}
The effective cross sections $\Sigma$ take into account the colour cross
sections and probabilities, as previously discussed. They are
\begin{eqnarray}
\Sigma_A(\tilde{z}_A,\vec{s}) & = & \!\! \sum_{j=S,O,D} 
\sigma^{in}_{\psi\,N_j}\ P^A_{j}(\tilde{z}_A,\vec{s})\,,\\
\Sigma_B(\tilde{z}_B,\vec{b}-\vec{s}) & = & \!\! \sum_{j=S,O,D} 
\sigma^{in}_{\psi\,N_j}\ P^B_{j}(\tilde{z}_B,\vec{b}-\vec{s})
\end{eqnarray}
In the above expressions the values $\tilde{z}$ correspond
to the $\psi$ absorption points as in the illustration shown in the left 
side of Figure \ref{TWO}. With a simple geometrical construction one obtains
\begin{equation}
\tilde{z}_A = z_A - (z'_B - z_B)\frac{1 - v_\psi}{1 + v_\psi}
\ \ \ \ \mbox{and}\ \ \ \ \tilde{z}_B = z_B - (z'_A - z_A)
\frac{1 + v_\psi}{1 - v_\psi}\,.
\end{equation}
The velocity $v_\psi$ of the meson is related to the measured value of
$x_F = p_\psi/p_{max} \simeq 0.15$ and to the centre of mass energy 
$\sqrt{s_{NN}}$ of the NN collisions. One has
$v_\psi = \sqrt{x_F^2s_{NN} / (4m_\psi^2 + x_F^2s_{NN})} \simeq 0.4$. 

The $\psi$ cross section expressed by eq.$\,$(\ref{NUCCROSS}) can now be 
evaluated. Its ratio with the Drell-Yan cross section is obtained in a 
conventional manner and is converted into a $E_T$-dependent function by fixing
the scale to the number of participants, in order to describe the minimum bias
data as measured by the NA50 experiment. No spread in the $E_T(\vec{b})$
correlation is taken into account for simplicity. The calculated ratio
is compared with the standard Glauber approach and to the data as shown in 
the right side of Figure \ref{TWO}. The result exhibits a 
stronger suppression when compared to the usual Glauber
calculation. Although several improvements are under investigation, the effect
is clear and cannot be neglected in future work. Among the aforementioned 
improvements is the inclusion of a retardation effect for the so far sudden 
switch of colour in NN collisions. This becomes relevant at large impact
parameters, where there are only few collisions, implying that the present
result overestimates the suppression at small transverse energy. Another 
improvement which works in the same direction consists in accounting for
important formation time effects \cite{HHK}. All this is presently under 
careful study.

\vspace{0.2cm}

I warmly thank J\"org H\"ufner and Boris Kopeliovich for the very pleasant 
and stimulating collaboration.

\end{document}